\documentclass{iopart}
\usepackage{iopams}
\usepackage{graphicx}
\usepackage[ansinew]{inputenc}
\usepackage{units}
\begin{document}

\title[Free space quantum key distribution with coherent polarization states]{Feasibility of free space quantum key distribution with coherent polarization states}

\author{D~Elser$^{1,2}$, T~Bartley$^{1,3}$, B~Heim$^{1,2}$, Ch~Wittmann$^{1,2}$, D~Sych$^{1,2}$ and G~Leuchs$^{1,2}$}

\address{$^1$ Institute of Optics, Information and Photonics, University of Erlangen-Nuremberg, Staudtstr.~7/B2, 91058~Erlangen, Germany}

\address{$^2$ Max Planck Institute for the Science of Light, Günther-Scharowsky-Str.~1, Building~24, 91058~Erlangen, Germany}

\address{$^3$ Physics Department, Blackett Laboratory, Imperial College, London SW7 2BZ, United Kingdom}

\ead{Dominique.Elser@mpl.mpg.de}

\begin{abstract}
We demonstrate for the first time the feasibility of free space quantum key distribution with continuous variables under real atmospheric conditions.
More specifically, we transmit coherent polarization states over a \unit[100]{m} free space channel on the roof of our institute's building.
In our scheme, signal and local oscillator are combined in a single spatial mode which auto-compensates atmospheric fluctuations and results in an excellent interference.
Furthermore, the local oscillator acts as spatial and spectral filter thus allowing unrestrained daylight operation.
\end{abstract}

\pacs{03.67.Dd, 42.25.Ja, 42.68.-w}

\submitto{\NJP}



\section{Introduction}

Quantum key distribution (QKD)~\cite{Gisin03,Scarani08} is the process of establishing a secret shared key between two parties, traditionally named Alice and Bob. The security is based on the laws of quantum mechanics, in contrast to classical schemes, where security relies on the unproven lack of efficient mathematical algorithms. A QKD system typically comprises a quantum channel and an authenticated classical channel. Alice and Bob initially exchange quantum states over the quantum channel which might be completely under the control of an eavesdropper (called Eve). In a second step, Alice and Bob use the classical channel to distil a secret key from measurement results taken during the first step. Eve may listen to this classical information but not modify it. In our experiment, we focus on the first step, more precisely on the generation, transmission and measurement of coherent polarization states providing the raw data for QKD. Our main focus is on the characterization of a real-world free space channel with regard to continuous-variable QKD.

\subsection{Free space quantum communication}

In classical telecommunication, free space optics (FSO)~\cite{Majumdar08} can help to solve the ``last mile'' problem: connecting end users to network nodes where installing optical fibres is often prohibitively time-consuming and expensive. Furthermore, FSO is utilized for satellite-to-ground and inter-satellite communication. In the domain of QKD, FSO offers an additional benefit: since fibre losses limit the maximum link range\footnote{This limitation could be overcome by using quantum repeaters which, however, are not yet available.}, FSO via satellite relays is currently the only feasible way to accomplish QKD over large distances.

The first demonstration of free space QKD over an atmospheric channel outside the laboratory was performed in 1996~\cite{Jacobs96}. Since then, several prepare-and-measure~\cite{Rarity01,Kurtsiefer02,Hughes02,Bienfang04,Schmitt-Manderbach07} and entanglement-based~\cite{Peng05,Ursin07,Ling08,Erven08} systems have been implemented. Currently, the world record distance is \unit[144]{km}~\cite{Schmitt-Manderbach07,Ursin07} and satellite QKD is in the early phases of development~\cite{Villoresi08,Perdigues08}. A common feature of all aforementioned systems is the use of single-photon detectors which, however, are impaired already by low background light intensities. Background light might stem from natural sources like the sun or the moon as well as from artificial illuminants such as street lamps~\cite{Rarity01}. In order to reduce this background noise, temporal, spectral and/or spatial filtering is employed~\cite{Miao05}. Despite this, performance in daylight is still degraded compared to night operation~\cite{Hughes02}.

In our system, we use an alternative approach: with the help of a bright local oscillator (LO) we perform homodyne measurements on weak coherent states. The primary task of the LO is to make the weak signal detectable by standard PIN photodiodes. Interestingly, apart from enabling homodyne measurements, the LO fulfils additional functions in our free space system:
\begin{description}
	\item[Spatial filtering:]Only photons which are spatially mode-matched to the LO produce a significant signal on the detector. The LO thus acts as a perfect single-mode spatial filter. Additional filtering (e.g.\ by pinholes or glass fibres) as used in single photon experiments is not required.	
	\item[Spectral filtering:]Our detection bandwidth can be precisely adjusted by electronic filtering of the homodyne signal. Background light outside this bandwidth does not perturb the measurement. Interference filters, commonly used in single photon experiments, exhibit orders of magnitude larger bandwidths and are lossy~\cite{Miao05}. Atomic filters~\cite{Shan06} perform better but still add complexity to the experiment.
	\item[Spatial tracking:]In our setup, the LO propagates in the same spatial mode as the signal. Thus spatial beam jitter and distortions can easily be monitored in order to compensate for them; no additional beacon beam is needed.
	\item[Timing generation:]Atmospherically induced time jitter can be determined from a pulsed LO. Thus the LO could fulfil the same task as the timing pulses in e.g.~\cite{Hughes02}.
\end{description}
For a homodyne detection, a good interference of signal and LO is crucial. Stabilizing this interference would be a problem if, as usual, signal and LO were propagating as two separate beams. In our setup, however, we use polarization states which allow for co-propagation of signal and LO in one single beam (see section~\ref{sec:Homodyne_Stokes_Measurement}). Thus the interference is intrinsically excellent which results in a high detection efficiency. Furthermore, phase fluctuations in the channel are auto-compensated.

\subsection{Continuous-variable quantum key distribution}

The well-known BB84 protocol~\cite{Bennett84} makes use of the non-orthogonality of single-photon polarization states ({\it discrete variable}). The implementations of BB84, however, mostly approximate single photons by attenuated coherent states which can be conveniently produced. In 1992, Bennett realized that, due to their inherent non-orthogonality, coherent states ({\it continuous variable}) can be directly used for QKD~\cite{Bennett92}. In the original B92 protocol, the detection was still performed by discrete single photon detectors. Later, continuous-variable protocols also using continuous homodyne detection were proposed~\cite{Ralph99}. PIN photodiodes used in homodyne detectors offer higher quantum efficiencies than single-photon avalanche photodiodes.

In homodyne QKD systems, signal and local oscillator (LO) typically propagate in the same channel. Hence the bright LO has to be multiplexed with the signal which otherwise would be masked.
In fibre-based experiments, temporal multiplexing by pulsing the LO~\cite{Lodewyck07,Legre06,Wittmann09} or spatial multiplexing by using two separate fibres~\cite{Lodewyck05} has been employed. Recently, combined polarization and frequency multiplexing in fibres has been demonstrated~\cite{Qi07}. In laboratory free space QKD systems, spatial~\cite{Hirano03,Lance05} or polarization~\cite{Lorenz04} multiplexing has been implemented. Using polarization multiplexing, signal and LO propagate in one spatial mode, such that no spatial interference stabilization is needed at the homodyne detector. This feature is particularly advantageous for atmospheric channels which are subject to spatial beam jitter.

In our continuous-variable QKD system, the following properties of a free space channel have to be considered:
\begin{description}
\item[Attenuation:]In security analysis, channel losses are attributed to Eve who can split off a part of each signal state and perform measurements on it. At channel losses of more than 50\%, Eve obtains more information about the signal than Bob. Although this problem can be circumvented by postselection~\cite{Silberhorn02} or reverse reconciliation~\cite{Grosshans03}, a low-loss channel is desirable since it allows for higher key rates.
In clear weather conditions, atmospheric attenuation in the transmission windows (e.g.\ between \unit[780]{nm} and \unit[850]{nm}) is indeed low ($<$\unit[0.1]{dB/km})~\cite{Scarani08}.

\item[Excess noise:]In discrete-variable systems, the polarization angle between two different signals is large (e.g.\ 22.5$^\circ$ in the BB84 protocol), therefore any birefringent effects of the atmosphere are negligible~\cite{Toyoshima08}. In our continuous-variable system, on the other hand, the polarization tilt between different signals is very small. Although atmospheric depolarization effects are small~\cite{Boyer76,Crosignani88}
they could cause significant excess noise (noise in excess of quantum noise resulting from the Heisenberg uncertainty principle). Excess noise is in principal assumed to stem from Eve's actions and therefore if it is too high security is compromised~\cite{Rigas06,Heid07,Zhao09}.

\item[Spatial beam jitter:]Atmospheric turbulence leads to fluctuations of the beam position at the receiver. Although this effect is not a fundamental limitation, it might lead to additional detection losses.
The receiver's optical elements should therefore be designed to collect all the incoming light. This can be done by choosing appropriate aperture diameters and/or by actively stabilizing the beam direction. 
In situations where this is technically not feasible, protocols such as quantum filtering~\cite{Wittmann08} can be employed.

\end{description}

\subsection{Homodyne detection of coherent polarization states}
\label{sec:Homodyne_Stokes_Measurement}

Polarization multiplexing of signal and LO~\cite{Lorenz04,Vidiella-Barranco06,Lorenz06} can be conveniently described in terms of Stokes operators~\cite{Korolkova02}. These operators are the quantum analogue to the classical Stokes parameters~\cite{Stokes1852} and read in our notation:
\begin{center}
	\includegraphics{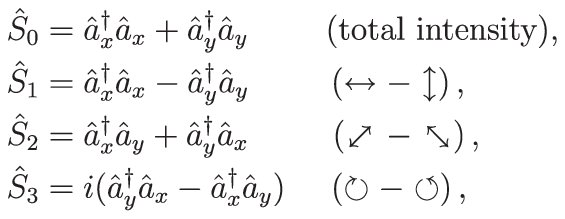}
\end{center}
in terms of the creation and annihilation operators $\hat{a}^{\dagger}$ and $\hat{a}$. The arrows in brackets display the operational definitions of the Stokes parameters as intensity differences of polarization types.

\begin{figure}
	\centering
		\includegraphics[width=0.50\textwidth]{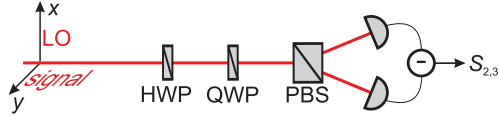}
	\caption{Stokes parameter measurement with half-wave plate (HWP), quarter-wave plate (QWP) and polarizing beam splitter (PBS). In our QKD experiment, the bright local oscillator is linearly polarized along the $x$-direction. Therefore, its only non-zero component is $S_1$. The $y$-polarized quantum signal is contained in the $S_2$ component. After appropriate adjustment of the wave plates, the homodyne signal appears in the difference between the currents of the two PIN photodiodes.}
	\label{fig:Stokes_Measurement}
\end{figure}

In our experiment, the photon number in the $x$-mode is much larger than in the $y$-mode. In this situation, we can describe a Stokes measurement of $S_{2/3}$ as homodyne detection of $\hat{a}_{y}$ with the local oscillator in $\hat{a}_{x}$ (see figure~\ref{fig:Stokes_Measurement}). This configuration results in the  uncertainty relation
\begin{equation}
\textrm{Var} ( \hat{S}_2 ) \cdot \textrm{Var} ( \hat{S}_3 ) \geq |\langle\hat{S}_1\rangle|^{2} .
\end{equation}
In the case of coherent states, the equality holds and the variances of $\hat{S}_2$ and $\hat{S}_3$ are equal. Excess noise leads to increased variances.

\section{Experimental setup}

\begin{figure}
	\centering
		\includegraphics[width=1.0\textwidth]{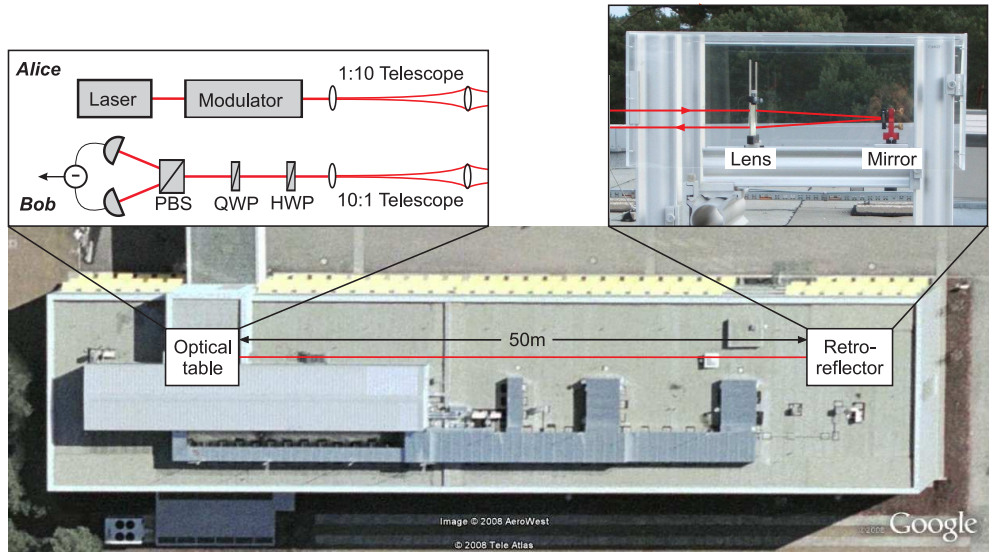}
	\caption{Experimental setup on the flat roof of our institute's building: Alice's laser emits a linearly polarized CW beam which later serves as a local oscillator for Bob's measurements. In terms of Stokes parameters, the local oscillator is $S_1$ polarized. Alice's magneto-optical modulator generates a weak signal in the $S_2$ component. The beam is expanded and sent to a retro-reflector at a distance of \unit[50]{m}. The retro-reflector is designed according to the cat's eye principle with a mirror placed at the focal length of a lens. Bob performs a homodyne Stokes measurement on the reflected beam.}
	\label{fig:Experimental_Setup}
\end{figure}

We transfer the principle of our earlier laboratory experiments~\cite{Lorenz04,Lorenz06} to a \unit[100]{m} atmospheric channel on the flat roof of our institute's building (see figure~\ref{fig:Experimental_Setup}).
In this proof-of-principle experiment, we place Alice's and Bob's station on one optical table inside the building and send the beam out and back using a retro-reflector. The reflected beam propagates exactly parallel but spatially displaced with respect to the outgoing one. The retro-reflector is designed according to the cat's eye principle, with a mirror placed at the focal length of an achromatic lens. A corner cube retro-reflector is not suitable for our polarization encoding since it translates spatial beam jitter into polarization fluctuations~\cite{Liu97}. Over the entire distance the beam propagates close to the roof surface. The temperature gradient of the roof and surrounding air thus provides us with appropriate real-world conditions for our investigations.

We use a grating-stabilized CW diode laser of wavelength \unit[809]{nm}, which lies within one of the atmospheric transmission windows~\cite{Scarani08}\footnote{Another optical atmospheric window is located around \unit[1550]{nm}. The larger beam diameters at the diffraction limit of these wavelengths, however, would demand larger optical elements.}.
After passing through a fibre mode-cleaner and a polarizing beam splitter, the laser beam is polarized along the $x$-direction. In other words, this light beam contains the $x$-polarized local oscillator in the $S_1$ component and vacuum in $S_2$ and $S_3$. By applying a current through the coil of the magneto-optical modulator (MOM) a magnetic field is generated. Via the Faraday effect, a magneto-optical crystal tilts the linear polarization of the light beam. This polarization tilt results in a weak $S_2$ component corresponding to an $y$-polarized signal in the same spatial mode as the local oscillator. The signal amplitude is taken out of the LO which, due to the weak modulation, remains essentially unaffected. 

\begin{figure}
	\centering
		\includegraphics[width=0.30\textwidth]{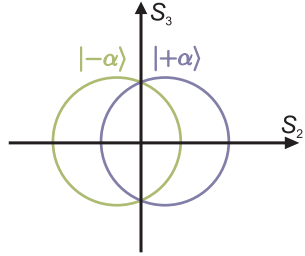}
	\caption{Contour plots of the two coherent states in our binary phase shift keying protocol. The amplitude $\alpha$ corresponds to the first moment of the Gaussian probability distributions. Due to the small amplitude the two states are nearly indistinguishable to Eve. By postselecting favourable measurement events, Bob gains an information advantage over Eve~\cite{Silberhorn02}.}
	\label{fig:2states}
\end{figure}

Using an arbitrary waveform generator, Alice randomly\footnote{Here we use pseudo-random numbers for experimental simplicity. To generate real random numbers, Alice could split off a part of the LO and perform a homodyne measurement on the vacuum state~\cite{Trifonov07}.} generates one of two coherent states $\left | + \alpha \right\rangle$ or $\left | - \alpha \right\rangle$ (see figure~\ref{fig:2states}). With this binary phase shift keying (BPSK) we use the smallest possible alphabet simplifying the analysis of statistical errors in a full security analysis. More complex alphabets along the $S_2$ axis could be easily generated by software modifications. For a modulation in the $S_3$ direction, an additional electro-optical modulator (EOM) could be incorporated into the setup~\cite{Lorenz04}. An EOM could also generate states on a circle around the origin~\cite{Sych09}.

Before transmission through the free space channel, the signal/LO beam is expanded by a 1/10 telescope such that its Rayleigh length corresponds to half the channel length. We thus obtain a collimated beam of diameter $\approx$\unit[1]{cm} whose waist we place onto the retro-reflector at a distance of \unit[50]{m}. Bob's 10/1 telescope, in turn, reduces the beam diameter to $\approx$\unit[1]{mm}. A receiving lens of diameter \unit[63]{mm} allows us to collect most of the spatially jittering beam without active stabilization.

In Bob's homodyne Stokes detection, the returning signal/LO beam is equally split into two parts by a Wollaston prism. Each beam is focused on a photodiode (Hamamatsu S3399) whose active area we chose to be larger than the beam size including beam jitter at the focal spot. The difference of the photocurrents from the two photodiodes is electronically amplified and processed by an analogue-to-digital converter. For an $S_2$ measurement, the detection basis is adjusted by a half-wave plate. Monitoring of the $S_3$ component~\cite{Lorenz06} is not performed in this study, but will be implemented in future experiments. Alice's arbitrary waveform generator and Bob's analogue-to-digital converter are currently embedded in the same computer. Thus synchronization is easily achieved by an electric cable which transmits trigger signals from Alice to Bob.

\section{Results}

We commence this section with an investigation of the noise behaviour of our newly developed magneto-optical modulator. Next, we present measurements of atmospheric polarization noise. From the absence of significant excess noise in both cases we deduce that our system is suitable for QKD operation. Finally, we demonstrate the transmission of quantum states over the atmospheric channel and provide calculations of the achievable key rate.

\subsection{Noise behaviour of the magneto-optic modulator}
\label{sec:MOM}

The bandwidth of our magneto-optical modulator (MOM) is limited by the inductance of the coil which generates the magnetic field. In the new version, the size of this coil has been decreased which enables us to operate the modulator at \unit[1]{MHz}.
When characterizing the MOM, we detect the modulated beam directly after modulation. We are therefore able to investigate the noise behaviour of the MOM separately from the free space channel.

A signal state is generated by applying a predefined positive or negative voltage to the MOM driver for \unit[400]{ns}. After each signal pulse the modulation voltage is switched to zero for \unit[600]{ns} to enable the modulator to return to its zero position. During this time, the signal is in the vacuum state. This vacuum reference is also needed for calibration since the polarization in the setup drifts slowly in time. We determine the vacuum level by calculating the mean value of 100 vacuum measurements neighbouring each signal pulse. This value is then subtracted from each signal measurement. At a constant vacuum level, an increased number of calibration pulses allows for a more precise calibration. In practice, however, this number is limited by slow polarization drifts as well as by laser excess noise at low frequencies.

\begin{figure}
	\centering
		\includegraphics[width=0.55\textwidth]{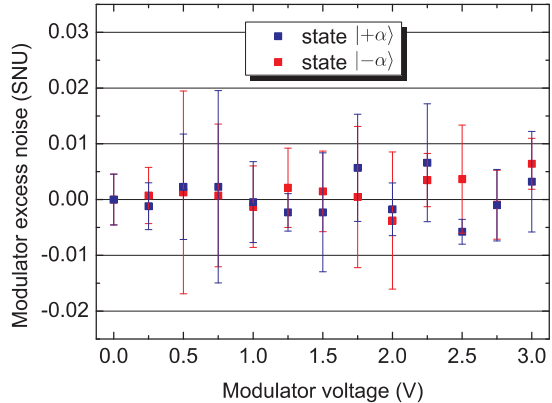}
	\caption{Excess noise (in shot noise units, SNU) introduced by the modulation of $S_2$. The modulation voltages are proportional to the amplitudes $\alpha$ of the coherent states (\unit[3.0]{V} corresponds to $\alpha=0.21$). The noise has been determined separately for negative and positive modulations. Within the measuring accuracy we find that no significant excess noise is caused by modulation using the MOM.}
	\label{fig:Modulation_Excess_Noise}
\end{figure}

We determine the excess noise of a signal state by comparing its variance to the variance of the vacuum state (shot noise). The variance of the vacuum state is normalized to unity.  Figure~\ref{fig:Modulation_Excess_Noise} shows the excess noise introduced by the modulation for different signal amplitudes. We see that, within the measuring accuracy, the modulation does not generate a significant amount of excess noise.

\subsection{Characterization of atmospheric polarization noise}

We measure atmospheric polarization noise by recording the noise of an unmodulated beam on a spectrum analyzer. We calibrate to shot noise by comparing the spectra of $\textrm{Var} ( \hat{S}_2 )$ before and after transmission through the free space channel. In both cases, the optical power at the detector is \unit[0.5]{mW}$\pm 3$\%. The measuring accuracy of the optical power is limited due to spatial beam jitter: the beam wanders across regions of slightly different quantum efficiency on the power meter. The measuring accuracy of the optical power $S_0$ translates to an inaccuracy in the noise measurement via the expression
\begin{equation}
|\langle\hat{S}_0\rangle| \approx |\langle\hat{S}_1\rangle| = \textrm{Var} ( \hat{S}_2 )
\end{equation}
which holds for a bright $S_1$ polarized beam at the quantum noise limit.

\begin{figure}
	\centering
		\includegraphics[width=0.70\textwidth]{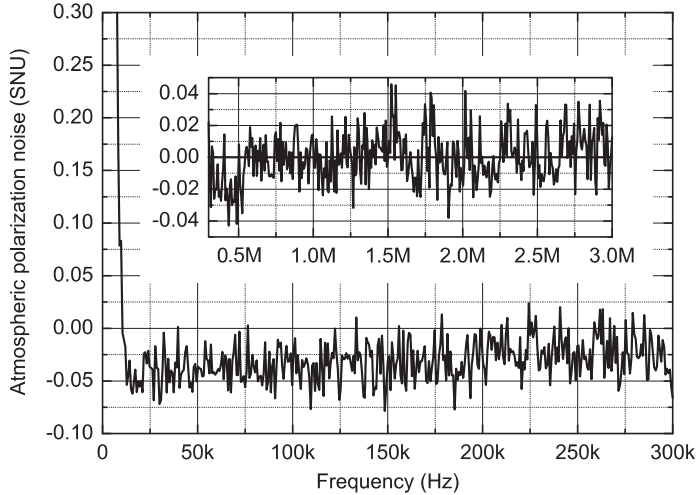}
	\caption{The determination of atmospheric polarization noise has been performed in two separate measurements from $0$ to \unit[300]{kHz} and from \unit[300]{kHz} to \unit [3]{MHz} (inset). Within the measuring accuracy of 3\% we find no significant excess noise for frequencies above \unit[10]{kHz}. We verified that background light at day or at night does not add noise to our measurements.}
	\label{fig:Atmospheric_Excess_Noise}
\end{figure}

As displayed in figure~\ref{fig:Atmospheric_Excess_Noise}, no significant excess noise is present above \unit[10]{kHz}. The noise below \unit[10]{kHz} amounts to less than 3 shot noise units which, due to the large magnification of the plot, result in an abrupt drop. This low-frequency noise stems at least partly from spatial beam jitter which causes the beam to wander across regions of slightly different quantum efficiency on the photodiodes. 
Our QKD signal at higher frequencies is not affected by the low-frequency noise. In a homodyne signal detection, however, the low-frequency deviations lead to an unbalance. In our measurements, this unbalance was typically around 1\% and never exceded 2\%. A calculation following~\cite{Lorenz05} reveals its impact on the quantum states: Firstly, since we adjust our laser to operate at the shot noise limit, the unbalance results in excess noise of less than 1\% which is lower than the detector's electronic noise. Secondly, the detection efficiency is varying by less than 1\%. This effect might give an eavesdropper an, albeit small, possibility to gain information~\cite{Fung09} and thus should be considered in future security analyses. As already mentioned in section~\ref{sec:MOM}, we have to choose the number of vacuum calibration pulses such that their level does not change due to low-frequency deviations. 

\subsection{Quantum signal measurements and estimation of secure key rate}

To obtain the optimal key rate, the signal amplitude $\alpha$ has to be adapted to the losses in the QKD system. We perform an optimization of $\alpha$ assuming postselection~\cite{Silberhorn02} and realistic two-way error correction procedures (\textsc{cascade}~\cite{Brassard94}). Intuitively, the existence of an optimum can be explained as follows:
For too small amplitudes, Bob obtains low information due to the high error rate. In the case of too large amplitudes, on the other hand, Eve's information is also large. To gain an information advantage, Bob must apply a high postselection threshold thus discarding many measurements.

\begin{figure}
\centering
\includegraphics[width=0.55\textwidth]{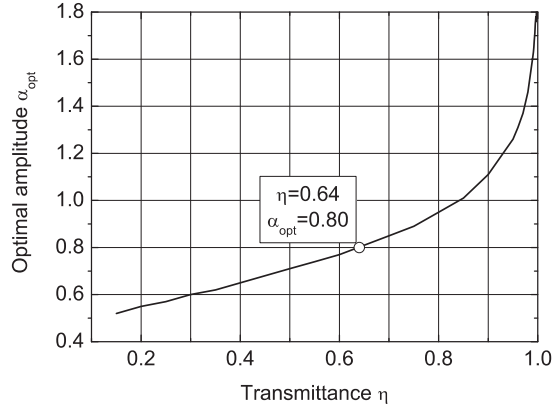}
\caption{Calculation of the optimal signal amplitude $\alpha_\textrm{opt}$ as a function of the transmittance~$\eta$. For the overall transmittance $\eta = 0.64$ in our experiment, we find the optimal amplitude $\alpha_\textrm{opt} = 0.80$.}
\label{fig:alpha_opt}
\end{figure}

Assuming a noiseless channel, the expected key rate $G$ is equal to the difference between Bob's and Eve's information~\cite{Heid06}:  
\begin{equation}
G=\int_{0}^{\infty}\left(1-f[e] H[e]-S[\hat\rho_E]\right)p(\beta) \textrm{d} \beta,
\label{eq:G}
\end{equation}
where $f[e]$ is the efficiency of \textsc{cascade}, $e$ is Bob's error rate, $H$ is the standard Shannon entropy, $p(\beta)$ is the probability of measuring the value $\beta$, and $S[\hat\rho_E]$ is the von Neumann entropy of Eve's density matrix~$\hat\rho_E$.
We apply postselection by accepting only positive contributions to the integral in~(\ref{eq:G}).
The exact value of the optimal amplitude $\alpha_\textrm{opt}$ is calculated numerically by maximizing the key rate for a given transmittance~$\eta$ (see figure~\ref{fig:alpha_opt}). The higher the transmittance, the less information can be potentially obtained by Eve, and the higher the optimal amplitude.

The transmittance of our \unit[100]{m} free space channel is $\eta_\textrm{ch} = 0.77$, with losses originating mainly from the retro-reflector.
In a conservative calculation we also attribute detection losses $(1-\eta_\textrm{det}) = 0.17$ to Eve. For the overall transmittance $\eta = \eta_\textrm{ch} \cdot \eta_\textrm{det} = 0.64 $, the optimal amplitude is $\alpha_\textrm{opt} = 0.80$.

\begin{figure}
	\centering
		\includegraphics[width=0.7\textwidth]{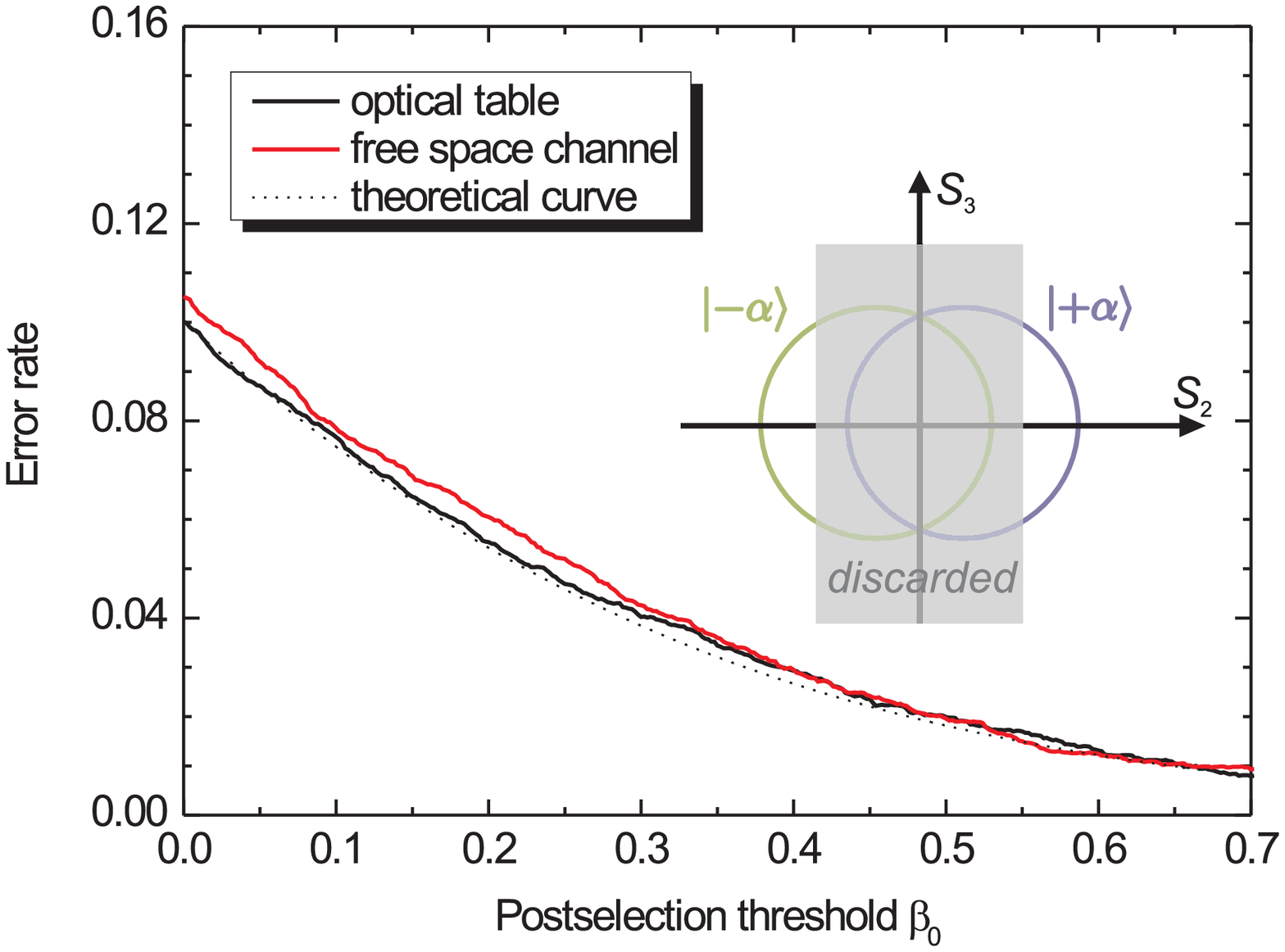}
	\caption{Error rates after postselection for coherent states with amplitude $\alpha = 0.80$. Increasing the postselection threshold (see inset) discards more ambiguous states and thus reduces the error rate. The measured error rates with and without the free space channel are equal within the measuring accuracy. This indicates that the quantum states are not affected by real atmospheric conditions. The measurements follow the theoretical curve $\frac{\textrm{erfc}[\sqrt{2} (\beta_0 + \sqrt{\eta} \alpha) ]}{2 P (\beta_0, \sqrt{\eta} \alpha)}$, where $P (\beta_0, \sqrt{\eta} \alpha)$ is the acceptance probability of the postselection~\cite{Namiki03}.}
	\label{fig:Error_Rates}
\end{figure}

At the moment, technical constraints prevent us from reaching the optimal amplitude at a pulse rate of \unit[1]{MHz}. For this reason, we choose a pulse rate of \unit[100]{kHz} in the following. 
Measurement results for the amplitude $\alpha_\textrm{opt} = 0.80$ are shown in figure~\ref{fig:Error_Rates} where we plot the error rates of measurements with and without free space channel. For better comparison, we adjust the detected optical power to be equal in both cases: any attenuation in the channel is effectively factored out. Thus the measured quantum states differ only with respect to their excess noise. Excess noise from the free space channel would increase the error rate. Within the measuring accuracy, however, we find no increased error rate which confirms that atmospheric excess noise is insignificant.

\begin{figure}
\centering
\includegraphics[width=0.55\textwidth]{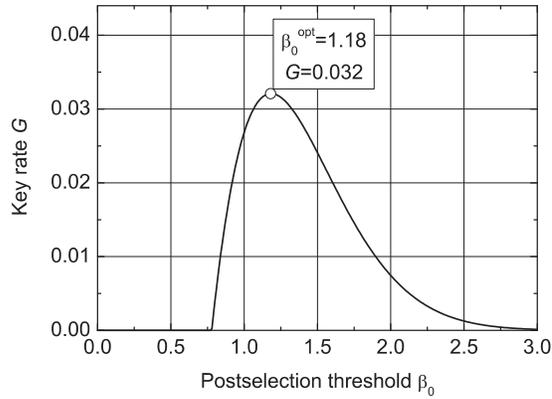}
\caption{Calculation of the key rate $G$ for the transmittance $\eta=0.64$ and the signal amplitude $\alpha=0.80$. At the optimal postselection threshold $\beta_0^{\textrm{opt}}=1.18$, the key rate is $G=0.032$.}
\label{fig:Keyrate}
\end{figure}

The optimal postselection threshold $\beta_0^{\textrm{opt}}$ is found by solving the equation $I_\textrm{AB}=I_\textrm{AE}$, where $I_\textrm{AB}$ is the mutual information between Alice and Bob, and $I_\textrm{AE}$ between Alice and Eve. This means Bob accepts only those signals which positively contribute to the total key. The key rate $G$ as a function of the postselection threshold $\beta_0$ is shown in figure~\ref{fig:Keyrate}. We can see that the postselection threshold has to be set correctly, otherwise the key rate is reduced.
At the optimal postselection threshold $\beta_0^{\textrm{opt}}=1.18$, the key rate would be \unit[3.2]{kbit/s} at a pulse rate of \unit[100]{kHz}.

\section{Conclusions and Outlook}
We have demonstrated the low-noise transmission of coherent polarization states over a real-world atmospheric channel. Our results indicate that our system is suitable for establishing a QKD link in urban environments in daylight.
To the best of our knowledge, this experiment is the first continuous-variable quantum communication under real atmospheric conditions. 

In future work, we want to extend our system to a point-to-point link in an urban environment. To synchronize Alice's and Bob's stations, we plan to interrupt the LO (and therefore also the signal) in regular time intervals. Switching on the LO will mark the beginning of a signal frame each containing about 1000 signal states. To avoid optical losses, we will choose the telescopes' diameter larger than the beam size including beam wander. Drifts in the telescopes' pointing direction could be compensated by an active pointing control~\cite{Weier03} with the LO as control signal.

Furthermore, we intend to increase the pulse modulation rate above \unit[10]{MHz}. For this frequency range, signal generators as well as digital-to-analogue converters are commercially available and detectors with low electronic noise have been built in our laboratory. Thus the limitation in pulse rate is only due to our magneto-optical modulator. However, the bandwidth of a magneto-optical modulator can in principle reach the GHz range (see~\cite{Irvine03} and references therein). Electro-optical modulators, on the other hand, are commercially available and could be used in our experiment as well~\cite{Lorenz04}.

Another interesting field of study is analyzing attacks on the implementation of our system. An eavesdropper could, for example, gain advantage by manipulating the local oscillator~\cite{Haeseler08} or by artificially tilting the beam to modify the efficiency of Bob's detectors~\cite{Fung09}.

\ack
We thank N~Lütkenhaus for very fruitful discussions.
D~Sych acknowledges the Alexander von Humboldt Foundation for a fellowship.
This work was supported by the EU project SECOQC.

\section*{References}
\bibliographystyle{iopart-num}
\bibliography{Elser_Free_space_CV_QKD}

\end{document}